\documentclass[hyper,12pt]{JHEP3}
\usepackage{epsfig}
\usepackage{amsmath}
\usepackage{amssymb}
\usepackage{epsf}
\usepackage{graphicx}
\usepackage{cite}
\usepackage{graphics}
\usepackage{epstopdf}

\newcommand\fverb{\setbox\pippobox=\hbox\bgroup\verb} \newcommand\fverbdo{\egroup\medskip\noindent% \fbox{\unhbox\pippobox}\
} \newcommand\fverbit{\egroup\item[\fbox{\unhbox\pippobox}]}
\newbox\pippobox

\newcommand{\nn}{\nonumber} \newcommand{\beq} {\begin{equation}}
\newcommand{\eeq} {\end{equation}} \newcommand{\beqa}
{\begin{eqnarray}} \newcommand{\eeqa} {\end{eqnarray}}

\newcommand{\ie}{{\it i.e.}}  
  \newcommand{\etal}{{\it et al.}}

  \newcommand{\morder}[1]{{\cal O}\left(#1 \right)}
\newcommand{\eq}[1]{(\ref{#1})}

\newcommand{\lsim}{\lesssim} \newcommand{\gsim}{\gtrsim}
 \newcommand{\im}{{\,\rm Im \,}}

\title{\center{Energy Loss of a Heavy Quark \\
Produced in a Finite Size Medium}}

\author{Pol-Bernard Gossiaux, St\'ephane Peign\'e\footnote{On leave of absence from LAPTH,
CNRS, UMR 5108, Universit\'e de Savoie, B.P. 110, F-74941
Annecy-le-Vieux Cedex, France}, Christina Brandt, J\"org Aichelin\\ SUBATECH, UMR 6457, Universit\'e de Nantes \\ Ecole des
Mines de Nantes, IN2P3/CNRS. \\ 4 rue Alfred Kastler, 44307 Nantes
cedex 3, France \\ 
E-mail: \email{gossiaux@subatech.in2p3.fr}, \email{peigne@subatech.in2p3.fr}, 
\email{christina.brandt@uni-rostock.de}, \email{aichelin@subatech.in2p3.fr}
}

\preprint{SUBATECH 2006-04 \\ LAPTH-1158/06}

\abstract{We study the medium-induced energy loss $-\Delta E_0(L_p)$ suffered by a heavy quark 
produced at initial time in a quark-gluon plasma, and escaping the plasma after travelling the 
distance $L_p$. The heavy quark is treated classically, and within the same framework $-\Delta E_0(L_p)$ consistently includes:
the loss from standard collisional processes, initial bremsstrahlung due to the sudden 
acceleration of the quark, and transition radiation. The radiative loss {\it induced by rescatterings} $-\Delta E_{rad}(L_p)$
is not included in our study. For a ultrarelativistic heavy quark with 
momentum $p \gsim 10 \ {\rm GeV}$, and for a finite plasma with $L_p \lsim 5 \ {\rm fm}$, the loss $-\Delta E_0(L_p)$ 
is strongly suppressed compared to the stationary collisional contribution $-\Delta E_{coll}(L_p) \propto L_p$. 
Our results support that $-\Delta E_{rad}$ is the dominant contribution to the heavy quark energy loss 
(at least for $L_p \lsim 5 \ {\rm fm}$), as indeed assumed in most of jet-quenching analyses. 
However they might raise some question concerning the RHIC data on large $p_{\perp}$ electron spectra.}


\begin{document}

\section{Introduction}

The attenuation of jets - jet quenching - was initially suggested by Bjorken \cite{bj} as a possible
signature of the quark-gluon plasma (QGP). 
Since, the suppression of hadron spectra at large transverse momentum $p_{\perp}$, in ultrarelativistic heavy ion 
compared to proton-proton collisions, has been observed \cite{phenix,star} at the Relativistic Heavy 
Ion Collider (RHIC). Until recently this suppression (and other features of jet quenching) 
seemed consistent \cite{Wang:2003mm} with the energy loss $-\Delta E_{rad}$ of the parent parton due to gluon radiation 
{\it induced by its rescatterings} \cite{Erad,GLV} in the hot or dense medium created in the heavy ion collision. 

However, recent experimental data on the $p_{\perp}$-spectra of electrons arising from heavy flavour 
decays \cite{Adler:2005xv,Bielcik:2005wu}, when compared to theory \cite{Djordjevic:2005db}, 
suggest that heavy quarks lose more energy than $-\Delta E_{rad}$ only\footnote{It is stressed in Ref.~\cite{Armesto:2005mz} 
that this statement might be somewhat premature, 
since the theoretical calculation of heavy quark production suffers from large uncertainties already 
in proton-proton collisions, and also because the contributions to the electron spectra from charm and beauty 
are not yet separated experimentally.}. It was suggested in Ref.~\cite{Wicks:2005gt} that 
this `single electron puzzle' might be solved by the {\it collisional} energy loss,
previously neglected in the total parton energy loss. 
However, one should remember that other sources of medium-induced energy loss are usually neglected, namely 
the Ter-Mikayelian (TM) effect and transition radiation. Thus it would be incorrect to invoke collisional 
loss without studying the effect of the latter radiative contributions\footnote{We stress again that 
the TM effect and transition radiation are radiative contributions distinct from that 
induced by rescatterings $-\Delta E_{rad}$.}. 
Transition radiation and the TM effect have already been studied, for instance in 
Refs.~\cite{Zakharov:2002ik,Djordjevic:2005nh} and \cite{DG} respectively. In this paper we present 
a theoretical model allowing to treat consistently the {\it zeroth order} of the (heavy quark) energy loss in 
an opacity expansion \cite{GLV}, denoted as $-\Delta E_0$, 
which includes the TM effect, transition radiation, and collisional energy loss.
Our main result is shown in Fig.~\ref{DeltaEofLp-fig}. For $L_p \lsim 5 \ {\rm fm}$ we find that 
$-\Delta E_0$ is suppressed compared to the collisional loss $-\Delta E_{coll}$,
indicating that $-\Delta E_{rad}$ should be dominant for such values of the plasma size
(contrary to what is suggested in \cite{Wicks:2005gt}), and leaving the `single electron puzzle' unsolved. 

In section \ref{secinf} we present our theoretical model, which is inspired from that of Ref.~\cite{PGG}, in 
the case of an infinite medium. We assume the heavy quark to be produced in a hard subprocess {\it at initial time}, 
in a static QGP of high temperature $T$
and small coupling $g \ll 1$. The latter hypothesis implies the hierarchy $1/T \ll r_D \ll \lambda$, where $1/T$ is 
the average distance between two constituents of the QGP, $r_D=1/m_D \sim 1/(gT)$ is the Debye radius and 
$\lambda \sim 1/(g^2 T)$ is the mean free path of the heavy quark. Under these 
hypotheses, we can describe the QGP via 
its {\em collective} response to the current \cite{TG}, where the (longitudinal and transverse) dielectric 
functions are obtained from the gluon polarization tensor. 
In \cite{PGG} we studied the {\it mechanical work} $W$ undergone by the heavy quark when travelling in the
(infinite) plasma, and used the hard thermal loop (HTL) \cite{BI} approximation for the 
dielectric functions. In section \ref{secinf} we show that to a very good accuracy, the model of 
\cite{PGG} can be simplified by assuming an isotropic plasma described by the dielectric function \eq{model}.
In fact, in the ultrarelativistic limit $\gamma = E/M = 1/\sqrt{1-v^2} \to \infty$ we are considering ($M$, $E$ and $v$ 
being the heavy quark mass, energy and velocity), 
the model \eq{model} reproduces {\it exactly} the results of \cite{PGG} for the work $W$. 
At finite but large $\gamma \gg 1$, the differences between the HTL approximation and the simple model \eq{model} are
numerically negligible, as shown in Fig.~\ref{dplot}. 
We end section \ref{secinf} by a critical discussion. We indeed stress that 
in the academic case of an infinite QGP, the {\it mechanical work} $W$ calculated in \cite{PGG} {\it should not} be 
interpreted as an observable energy loss (contrary to what was done in \cite{PGG}). This is due to the fact that the 
quark asymptotic ($t = +\infty$) states are different in vacuum and in an infinite QGP \cite{pol}. 

In section \ref{secfin} we consider the more realistic case where the quark is produced in a hard subprocess 
in a {\it finite size} (but still static) QGP, and travels the distance $L_p$ before escaping the medium. 
In this case the ``asymptotic'' quark states are the same in vacuum and in the presence of the medium, 
and it is legitimate to interpret the work $-W$ as the (medium-induced) quark energy loss, 
provided one includes transition radiation. We thus study the heavy quark energy loss $-W \equiv -\Delta E(L, L_p)$, 
where $L$ is the total distance travelled by the (deconfined) 
quark, related to the quark hadronization time as $L=v t_\mathrm{had} \sim v \gamma /\Lambda_\mathrm{QCD}$. 
We assume the heavy quark to be relativistic
enough so that it hadronizes outside the medium, $t_\mathrm{had} \simeq L \gg L_p$. We also consider a high 
temperature QGP, with a Debye mass $m_D \sim gT \gg \Lambda_\mathrm{QCD}$. The hadronization time being 
larger\footnote{For a charm quark the assumption $t_\mathrm{had} \gg L_p$ should 
be reasonable in practical applications where $L_p \lsim 5\,{\rm fm}$ and $E \geq 10\,{\rm GeV}$.} than the other 
relevant length scales of the problem $L_p$ and $1/m_D$, the loss 
$-\Delta E(L=\infty, L_p) \equiv -\Delta E_0(L_p)$ will be the main quantity of
interest, affecting the final hadron (or decay electron) $p_{\perp}$-spectrum. 
The calculation of transition radiation fields in section \ref{secfin} is done using the simple model 
\eq{model} motivated in section \ref{secinf}, and our 
final result for the {\it zeroth order} energy loss $-\Delta E_0(L_p)$ is shown in Fig.~\ref{DeltaEofLp-fig}. 
 
In the present study we focus on the case of a heavy quark of mass $M$. 
Comparing $-\Delta E_0(L_p)$ to the stationary collisional energy loss $-\Delta E_{coll}(L_p) \propto L_p$ \cite{BT}, 
our estimate (see \eq{dprimeinf} and 
Fig.~\ref{DeltaEofLp-fig}) $(-\Delta E_{coll}(L_p)+\Delta E_0(L_p))/E = \sqrt{2}/3 \, C_F \alpha_s m_D /M \simeq 5 \% $, 
for a charm quark of mass $M = 1.5 {\rm GeV}$ and for $L_p > 4 {\rm fm}$, is numerically consistent with 
\cite{Djordjevic:2005nh,Djordjevic:2006tw}. 
However we find explicitly that the reduction of $-\Delta E_0(L_p)$ compared to 
$-\Delta E_{coll}(L_p)$ scales as $\gamma$ when $\gamma \to \infty$, which can be interpreted 
as an effective 'retardation' of the stationary regime (see Fig.~\ref{DeltaEofLp-fig}) scaling similarly.
We think it is important to extract such an asymptotic parametric behaviour in order to get a better 
insight on heavy quark energy loss, and also in view of applications to heavy ion collisions at higher
energies such as those planned at the Large Hadron Collider.  
We recall that our calculation includes, in a {\it unique} classical framework, 
transition radiation, initial bremsstrahlung
and collisional processes. The latter were not included in \cite{Zakharov:2002ik} and were calculated
in \cite{Djordjevic:2006tw} without the contribution from transition radiation (calculated separately 
in \cite{Djordjevic:2005nh}). Let us mention Ref.~\cite{Wang:2006qr}, where
an effect of interference between elastic and radiative amplitudes is studied. This effect is not included in our 
approach.

We stress that our results should be taken at the qualitative level only. Indeed, we work 
in an ideal framework, where the medium created in the heavy ion collision is assumed to be a static, equilibrated, high 
temperature QGP with small coupling $g \ll 1$. 
We also work in the fixed coupling approximation. In the case of a running $\alpha_s$ 
we expect the slope $(-dE/dx)_{coll}$ of the stationary collisional loss to be independent of $E$ 
when $E \to \infty$ \cite{Peshier:2006hi}, 
contrary to the result for fixed $\alpha_s$ \cite{TG,BT} which is logarithmic in $E$ (see \eq{linearlaw}). 
Despite this reduced energy dependence, $(-dE/dx)_{coll}$ with running $\alpha_s$ might be larger than in the 
fixed coupling approximation, because it behaves as  $(-dE/dx)_{coll} \propto \alpha_s$ instead of 
$\propto \alpha_s^2$ \cite{Peshier:2006hi,Peshier:2006mp}. We thus cannot exclude that a larger $(-dE/dx)_{coll}$ 
than usually assumed partly compensates the drastic suppression we find for $-\Delta E_0(L_p \lsim  5 {\rm fm})$. 
A consistent calculation of $-\Delta E_0(L_p)$ (and of the slope of its asymptote $(-dE/dx)_{coll}$) 
{\it with running $\alpha_s$} is needed to answer this question.

\section{Infinite medium: theoretical model and critical discussion}
\label{secinf}

In this section we first recall the model used in \cite{PGG} (which was directly adapted from
Ref.~\cite{TG}) to study the mechanical work $W$ undergone by a heavy quark produced at $t=0$ in an infinite QGP,
and travelling the distance $L$ in the plasma. We then simplify this model in order to focus
on its essential feature. 

We start from the expressions of the longitudinal and transverse (chromo-)electric fields 
in momentum space $K=(\omega, \vec{k})$,
\beq
\label{linappr}
\vec{E}_L^a = \frac{4\pi}{i\omega} \,  \frac{\vec{j}_L^a}{\epsilon_L} \ \ \ \ \ \ ;\ \ \ \ \ \ 
\vec{E}_T^a = \frac{4\pi}{i\omega} \, \frac{\vec{j}_T^a}{\epsilon_T - k^2/\omega^2} \ \ \ \ , 
\eeq 
which follow from Maxwell equations in a medium with dielectric functions 
$\epsilon_L$ and $\epsilon_T$, in the abelian approximation and within linear response theory.
For a heavy quark of color charge $q^a$ (with $q^a q^a = C_F \alpha_s$) produced at $t=0$, 
the classical {\it $3$-vector} current density $\vec{j}^a$ is of the form \cite{PGG}:
\beq
\label{jmom}
\vec{j}^a(K) = i q^a \frac{\vec{v}}{K.V + i\eta}  \ \ ,
\eeq
with longitudinal and transverse components given by 
$\vec{j}_{L} = (\vec{j}.\vec{k}/k^2)\vec{k}$ (where $k=|\vec{k}|$) and
$\vec{j}_{T} = \vec{j}-\vec{j}_{L}$. The quark $4$-velocity is denoted by $V=(1, \vec{v})$ and 
assumed to be constant in the following. Let us mention that we can easily form a {\it conserved} 
$4$-current whose vector component is precisely given by \eq{jmom}, for instance by assuming the 
fast heavy quark to be produced in conjunction with a static antiquark at $t=0$. 
This was done in Ref.~\cite{PGG}, where it was shown that the dominant effect (scaling as $\gamma$ when 
$\gamma \to \infty$, see \eq{dinf2} below) on $W$ arises from the fast quark contribution to the 
conserved current, given by \eq{jmom}. 
The precise form of the conserved current used in \cite{PGG} only affects the {\it longitudinal} 
contribution to the medium-induced energy loss, which receives a term corresponding to the 
difference of the dipole binding energies in medium and in vacuum. When $\gamma \gg 1$ this contribution is 
subleading \cite{PGG} compared to the 
dominant ({\it transverse}) contribution $\propto \gamma$. In the present 
study, where we focus on the leading effects scaling as $\gamma$, we can thus consider
\eq{jmom} as a relevant model for the current. 

In coordinate space the medium-induced electric field reads
\beq
\label{efield}
\vec{\cal{E}}^a(t, \vec{x}) = \int \frac{d^4 K}{(2\pi)^4} \, e^{-i (\omega t - \vec{k}.\vec{x})} 
\left[\vec{E}_L^a  + \vec{E}_T^a \right]_{\rm ind} \ , 
\eeq 
where the `induced' prescription corresponds to subtracting the vacuum $\epsilon_L=\epsilon_T=1$ 
contribution. The induced mechanical work $W$ of the electric force on the quark is given by 
\beq \label{eloss}
- W = -\vec{v} \cdot \int_0^{L/v} dt\, q^a \vec{\cal{E}}^a(t, \vec{v} t) 
= - q^a \vec{v} \cdot \int \frac{d^4 K}{(2\pi)^4} \int_0^{L/v} dt \, e^{-i K.V\, t} 
\left[ \vec{E}_L^a + \vec{E}_T^a \right]_{\rm ind} \ \ .
\eeq 

In Ref.\cite{PGG} the work $W$ is calculated by choosing for 
the dielectric functions $\epsilon_L$ and $\epsilon_T$ those of a high temperature QGP.
The latter have a rich analytical structure, which somewhat complicates the calculation of 
the energy loss. In par\-ti\-cu\-lar, the dielectric functions have a cut 
in the spacelike $|\omega| < k$ region (physically corresponding to Landau damping), responsible 
for the leading large $L$ behaviour,
\beq
-W(L) \mathop{\ \ \simeq \ \ }_{L \to \infty} - \Delta E_{coll}(L)+ \morder{L^0} 
\ \ \ {\rm where} \ \ \ - \Delta E_{coll}(L) \equiv \left(-\frac{dE}{dx} \right)_{coll} L  \ \ \ .
\label{largeL}
\eeq
We are interested in the difference, denoted as $d(L)$, between $-W(L)$ and the stationary law
for collisional loss $-\Delta E_{coll}(L)$, 
\beq
\label{ddef}
d(L) = -W(L) + \Delta E_{coll}(L) = -W(L) + \left(\frac{dE}{dx}\right)_{coll} L  \ \ \ .
\eeq
By definition the function $d(L)$ tends to a constant $d_\infty$ when $L \to \infty$.

It has been shown \cite{PGG} that for $\gamma \gg 1$, $d(L)$ arises 
dominantly from the transverse contribution and from the phase space region $k \sim \gamma m_D$ and $x=\omega/k \to 1$. 
In this limit the HTL \cite{BI} transverse gluon propagator takes a simple form,
\beq
\label{prop}
\Delta_T(\omega=k x, k) = \frac{-1}{\omega^2-k^2-\Pi_T(x)} \equiv \frac{-1}{\epsilon_T \, \omega^2-k^2} 
\mathop{\ \ \simeq \ \ }_{x \to 1} \frac{-1}{\omega^2-k^2-m^2}  \ \ ,
\eeq
where $m=m_D/\sqrt{2}$ is the asymptotic (transverse) gluon thermal mass and the retarded prescription
$\omega \to \omega + i\eta$ is implicit. Thus $d(L)$ can be simply modelled by assuming a medium with dielectric 
function
\beq
\label{model}
\epsilon(\omega) = 1 -\frac{m^2}{\omega^2} \ \ .
\eeq
All medium effects are encoded in the single parameter $m$. As we will shortly see, the model \eq{model} 
captures the exact leading order in $\gamma \gg 1$, and will greatly simplify the calculation of 
transition radiation in the next section, due to $\epsilon(\omega)$ being independent of $k$ - corresponding to the 
plasma response involving no spatial dispersion\footnote{The isotropy implied by \eq{model} 
obviously imposes $\epsilon_L = \epsilon_T = \epsilon(\omega)$.}. The main effects included in our calculation 
- initial bremsstrahlung and transition radiation -  will arise from the domain 
$\omega \simeq k \propto \gamma m$, justifying a posteriori using \eq{model}. We evaluate \eq{eloss} by 
performing the $\omega$-integral using Cauchy's theorem, and the function $d(L)$ by subtracting the leading term
(linear in $L$) when $L \to \infty$, 
\beqa
\label{dofl}
&& d(L) = \frac{C_F \alpha_s}{\pi^2} \int d^3\vec{k} 
\left\{ \vec{v}_T^{\,2} \frac{\sin^2{\left[(E_k - \vec{k}\cdot\vec{v}) L/(2v) \right]}}{(E_k - \vec{k}\cdot\vec{v})^2} \right.  \nn \\
&& \hskip 1.5cm  \left. +  \vec{v}_L^{\,2} \, 
\left[ \frac{\sin^2{\left[(m - \vec{k}\cdot\vec{v}) L/(2v) \right]}}{(m - \vec{k}\cdot\vec{v})^2} 
- \frac{\pi L}{2 v} \delta{(m - \vec{k}\cdot\vec{v})}\right]  \right\}_{\rm ind}   \ \ \ ,
\eeqa
where $E_k = \sqrt{k^2+m^2}$ and the 
`induced' prescription corresponds to subtracting the value of the integrand at $m=0$. In \eq{dofl} the second 
line corresponds to the longitudinal contribution, which will be subleading in the following. 

The explicit calculation of \eq{dofl} yields:
\beq
d_\infty \equiv \mathop{\rm lim}_{L \to \infty} d(L) 
= - 2 C_F \alpha_s m \left[ \frac{1}{\sqrt{1-v^2}} - \frac{\arcsin{v}}{v}  + \frac{\pi}{2 v} \right] \ \ , 
\label{dinf}
\eeq
where the last term in the brackets stands for the longitudinal contribution. 
In the $v \to 1$ limit, \eq{dinf} reproduces exactly the result of \cite{PGG} obtained within the HTL approximation, 
and arises dominantly from the transverse contribution\footnote{We note that the longitudinal
contribution given in the second line of \eq{dofl} is not the same as in Ref.\cite{PGG}. In particular it does not include
the induced binding energy of the dipole created at $t=0$. This is because instantaneous ($\omega = 0$) 
self-interactions are suppressed in the model \eq{model}. These details turn out to be irrelevant since the 
longitudinal contribution is subleading when $\gamma \gg 1$, both in Ref.~\cite{PGG} and in the present study.}:
\beq
d_\infty \mathop{\simeq}_{v \to 1}  - \sqrt{2} C_F \alpha_s m_D \, \gamma  \ \ .
\label{dinf2}
\eeq

\begin{figure}[t]
  \centering
\includegraphics[width=7cm]{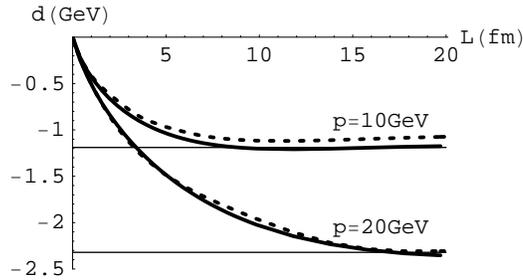}
  \caption[*]{The function $d(L)$ defined by \eq{ddef} and given by \eq{dofl} 
in the model \eq{model} (thick lines). We consider a charm quark of mass $M=1.5\,{\rm GeV}$, use
$\alpha_s = 0.2$, $m^2 = m_D^2/2 =  2\pi\alpha_s T^2(1+n_f/6) \simeq (0.32\,{\rm GeV})^2$ for 
$n_f=2$ flavours and a plasma temperature $T=0.25\,{\rm GeV}$. 
We show $d(L)$ for two values of the charm quark momentum, $p=10\,{\rm GeV}$ and $p=20\,{\rm GeV}$. The thin
straight lines give the values of $d_\infty$ (see \eq{dinf}) for the corresponding 
values of $p$. The (HTL) results of Ref.~\cite{PGG} for $d(L)$ are given by the dashed lines.}
\label{dplot} 
\end{figure}

The numerical evaluation of $d(L)$ is displayed in Fig.~\ref{dplot}, where it is also compared to 
the results of \cite{PGG} obtained with the HTL expressions of the 
dielectric functions. 
Since the results of Ref.~\cite{PGG} for $d(L)$ are very accurately reproduced 
with the simple model \eq{model}, we will use the latter model in section \ref{secfin} in order to derive 
the 'zeroth order' of the heavy quark energy loss $-\Delta E_{0}(L_p)$ in a finite QGP.  

Before doing that let us discuss (see also \cite{pol}) the incorrect interpretation which was done in \cite{PGG}. 
In \cite{PGG} the stationary collisional loss $-\Delta E_{coll}(L) \propto L$ \cite{BT} was added 
to $d(L)$, and the result {\it interpreted} as the quark {\it energy loss} $-\Delta E(L)$
in a infinite plasma. 
Due to the large negative value of $d_\infty$, a significant delay before the onset of the stationary (linear) behaviour
was observed, and designated as 'retardation effect'. 
As noted in \cite{PGG} this 'retardation effect' is not a genuine retardation of purely collisional loss, 
since various physical effects contribute to the apparent delay. First, part of the work done on the charge is 
due to the initial radiation induced by the sudden acceleration of the quark at $t=0$. 
Due to the difference between the gluon dispersion relations in medium and in 
vacuum, initial bremsstrahlung in QGP differs from that in
vacuum. This is the QCD equivalent of the Ter-Mikayelian (TM) effect, studied in \cite{DG}
and also properly identified in \cite{PGG}. In the limit $\gamma\gg 1$, half of the 
apparent `retardation' is actually due to the TM effect \cite{PGG}. 

As can already be seen in the case of a charge with $v \simeq 1$ produced in vacuum, 
initial radiation represents only {\em half} of the mechanical 
work done on the charge after its production. Using Poynting's theorem we can 
show that the other half is associated to the creation of the charge's proper 
field. Contrary to what was implicitly done in \cite{PGG}, this {\it self-energy} contribution should not be 
counted as `energy loss' as it is part of the charge asymptotic state. 
With an accurate definition of energy loss in terms of asymptotic states \cite{pol}, we 
find that the retardation time of {\it purely collisional} energy loss is quite small, 
$t_{\rm ret} \sim 1/m_D$, instead of $t_{\rm ret} \propto \gamma/m_D$ as argued in \cite{PGG}.

In the next section we will consider the more realistic situation, where the quark is produced 
in a finite size QGP. In this case the `asymptotic' (we assume the quark to hadronize long after 
escaping the medium) quark self-energy is the same as in vacuum. Interpreting (minus) the mechanical work on the quark 
as energy loss is then legitimate, provided transition radiation is taken into account. 
Let us stress here that our final result for $-\Delta E_{0}(L_p)$ in Fig.~\ref{DeltaEofLp-fig} 
displays an {\it effective} time delay $t_{\rm ret} \propto \gamma/m_D$ before the linear regime. 
This effective delay is not due to a retardation of collisional energy loss (at least at leading order in 
$\gamma \gg 1$), but arises from a non-compensation - already noted in \cite{Djordjevic:2005nh} - 
between the TM effect and transition radiation.

\section{Finite size medium: implementing transition radiation}
\label{secfin}

Here we study how the finite size $L_p$ of the medium affects the work $W$ done on the heavy quark. 
Instead of $W(L) \equiv W(L, L_p=\infty)$, we now consider the explicit dependence of 
$W(L, L_p)$ on $L_p$, and implement transition radiation induced by the discontinuity at $L=L_p$ between medium and 
vacuum.

We derive transition radiation along the lines of Ref.~\cite{LL}. We assume that the separation surface between 
medium and vacuum is a plane located at $z=L_p$, and that the heavy quark produced at $t=0$ and $z=0$ 
in the medium (with the associated current \eq{jmom}) travels along the $z$-axis (with $v^z= v >0$). 
We denote the medium and vacuum as media 1 and 2, with dielectric functions
$\epsilon_{1}(\omega)$ given by \eq{model} and $\epsilon_{2}(\omega)=1$.
(This is similar to the model used in \cite{Djordjevic:2005nh}). 
Transition radiation is obtained by adding to the fields \eq{linappr} in both 
media some (transverse) fields $\vec{E}_{T}^{\, '}$ solutions of the homogeneous (abelian-like) Maxwell equations. 
It is convenient \cite{LL} to express these fields in the mixed $(\omega,\vec{k}_\perp,z)$ representation, 
\beqa
\vec{E}_{T1}^{\, 'a}(\omega,\vec{k}_\perp,z) &=& 4\pi q^a \, h_1(\omega,k_\perp) \, e^{-i \sigma_1 (z-L_p)}
\left(k_\perp^2 \vec{e}_z + \sigma_1 \vec{k}_\perp\right) \nn \\
\vec{E}_{T2}^{\, 'a}(\omega,\vec{k}_\perp,z) &=& 4\pi q^a \, h_2(\omega,k_\perp) \, e^{i \sigma_2 (z-L_p)}
\left(k_\perp^2 \vec{e}_z - \sigma_2 \vec{k}_\perp\right) 
\label{homfields} \\ 
&& \hskip -0.5cm 
\sigma_i = \sigma_i(\omega,k_\perp) \equiv \sqrt{\epsilon_{i} \, \omega^2 - k_{\perp}^2} \ \ , \nn
\label{chidef}
\eeqa
where the subscript ``$\perp$'' denotes a vector component orthogonal to the unit vector $\vec{e}_z$ 
specifying the $z$-axis, the retarded prescription $\omega \to \omega + i \eta$ is implicit, and 
the square root is defined as having a cut along the {\it positive} real 
axis, \ie\ such that $\im \sqrt{z} \geq 0$ for all complex $z$.
One directly sees that the fields \eq{homfields} satisfy\footnote{Recall that the permittivity does not 
depend on $k_z$ in the model \eq{model}.} the (homogeneous) Gauss and d'Alembert's equations, which read 
$\partial_z E_z + i\vec{k}_\perp\cdot \vec{E}_\perp=0$ and 
$(\epsilon \, \omega^2 -k_\perp^2)\vec{E}+\partial_z^2 \vec{E}=\vec{0}$ in the mixed representation. 

With our definition of the square root we have 
\beqa
\sigma_i=
\left\{
\begin{array}{cc}
{\rm sign}(\omega) \sqrt{\epsilon_{i} \, \omega^2 - 
k_{\perp}^2}&\text{for $\epsilon_{i} \, \omega^2>k_\perp^2$} \ \ \ {\rm region\ (i)}\\[3mm]
i \sqrt{k_{\perp}^2-\epsilon_{i} \, \omega^2 }&\text{for $\epsilon_{i}\,
\omega^2<k_\perp^2$} \ \ \ {\rm region\ (ii)}
\end{array}
\right. \hskip 1cm  \ \ .
\eeqa
The electric field in coordinate space arises from either region (i) or (ii) 
in momentum space. The contributions from region (i) correspond to a superposition of plane waves 
propagating with decreasing $z$ (for $\vec{E}_{T1}^{\,'}$) or increasing $z$ (for $\vec{E}_{T2}^{\,'}$), 
in a direction specified by the wave vector $\vec{k} = (\vec{k}_\perp, -\sigma_1)$ or  $\vec{k} = (\vec{k}_\perp, \sigma_2)$. 
The absence of components $\propto e^{+i \sigma_1 z}$ in $\vec{E}_{T1}^{\,'}$ and 
$\propto e^{-i \sigma_2 z}$ in $\vec{E}_{T2}^{\,'}$ is dictated by the transition fields being 
created at the interface between medium ($z < L_p$) and vacuum ($z>L_p$), so that no wave comes from 
$z=-\infty$ in the medium or from $z=+\infty$ in the vacuum. 
The contributions from region (ii) correspond to surface waves which decay exponentially in the longitudinal 
distance. Including terms 
$\propto e^{+i \sigma_1 z}$ in $\vec{E}_{T1}^{\,'}$ and $\propto e^{-i \sigma_2 z}$ in $\vec{E}_{T2}^{\,'}$ would be 
unphysical as they diverge for $z\rightarrow -\infty$ and  $z\rightarrow +\infty$ respectively. 
We stress that both types of waves (plane waves and surface waves) are simultaneously included in our framework 
by defining the square root with a cut along the positive real axis (with the implicit retarded prescription 
$\omega \to \omega +i\eta$). 

The functions $h_1$ and $h_2$ in \eq{homfields} are obtained by requiring that the total 
(inhomogeneous) fields
\beqa
\vec{E}_1^a &=& \vec{E}_{L1}^a + \vec{E}_{T1}^a + \vec{E}_{T1}^{\, 'a}  \nn \\
\vec{E}_2^a &=& \vec{E}_{L2}^a + \vec{E}_{T2}^a + \vec{E}_{T2}^{\, 'a}   
\label{totalfields}
\eeqa
are consistent with the continuity of the components 
$\vec{E}_{\perp}(t,\vec{x}_{\perp}, z)$ and 
$\vec{D}^{z}(t, \vec{x}_{\perp}, z)$ at the transition surface $z=L_p$. 
We evaluate $h_1(\omega, k_\perp)$ and $h_2(\omega, k_\perp)$ explicitly in Appendix A. 
The results are given by \eq{h1h2app} and \eq{I1I2exp}. 

The work $-W(L, L_p)$ is given 
by an expression similar to \eq{eloss}, but we now have to specify whether the quark is in medium 1 or 
medium 2 (the vacuum) at time $t$:
\beq 
\label{elosstrans}
-W(L, L_p) = - q^a \vec{v} \cdot \int_0^{L/v} dt\, \left\{ \Theta(L_p -v t) \vec{{\cal E}}_1^a(t, \vec{v} t) 
+ \Theta(v t - L_p) \vec{{\cal E}}_2^a(t, \vec{v} t) \right\} \ \ .
\eeq 
The medium-induced fields are related to the total fields \eq{totalfields} which include the transition fields.
Instead of \eq{efield} we have
\beq
\label{efieldtrans}
\vec{{\cal E}}_i^a(t, \vec{x}) = \int \frac{d^4 K}{(2\pi)^4} \, e^{-i (\omega t - \vec{k}.\vec{x})} 
\left[\vec{E}_{Li}^a  + \vec{E}_{Ti}^a + \vec{E}_{Ti}^{\, 'a} \right]_{\rm ind} \ . 
\eeq 
Inserting \eq{efieldtrans} in \eq{elosstrans} and subtracting the leading term (linear in $L$) when $L \to \infty$ we get:
\beq
d(\hat{L})  - W'(L, L_p) \equiv d'(L,L_p) \hskip 1cm ;  \hskip 1cm \hat{L} \equiv {\rm Min}\{L, L_p\} \ \ ,
\label{twoterms}
\eeq
where $d(L)$ is given by \eq{dofl} and the contribution $-W'(L, L_p)$ arises from the transition fields:
\beqa
- W'(L, L_p) = - q^a \vec{v} \cdot \int \frac{d^4 K}{(2\pi)^4} \left\{ \int_0^{\hat{L}/v} dt\, e^{-i K.V t} \vec{E}_{T1}^{\, 'a}
\right. \ \ \ && \nn \\   \left. + \Theta(L - L_p) \int_{L_p/v}^{L/v} dt\, e^{-i K.V t} \vec{E}_{T2}^{\, 'a} \right\} && \ \ .
\eeqa
Using the mixed representation \eq{homfields} of the transition fields and integrating over time we obtain:
\beqa
-W'(L, L_p) = \frac{C_F \alpha_s}{2 \pi} \int_0^{\infty} dk_\perp^2 \, k_\perp^2 \int_{-\infty}^{\infty} d\omega 
\left\{ \frac{1-e^{-i(\frac{\omega}{v} +\sigma_1) \hat{L}}}{\frac{\omega}{v} +\sigma_1} \, i h_1(\omega, k_\perp) \,
e^{i \sigma_1 L_p} \right. && \nn \\
\left. + \Theta(L - L_p) \frac{e^{-i(\frac{\omega}{v} -\sigma_2) L_p}-e^{-i(\frac{\omega}{v} -\sigma_2) L}}{\frac{\omega}{v} -\sigma_2} \, 
i h_2(\omega, k_\perp) \, e^{- i \sigma_2 L_p}\right\} && \ . \nn \\
\label{mastereq}
\eeqa
 
\begin{figure}[t]
\centering
\includegraphics[width=7cm]{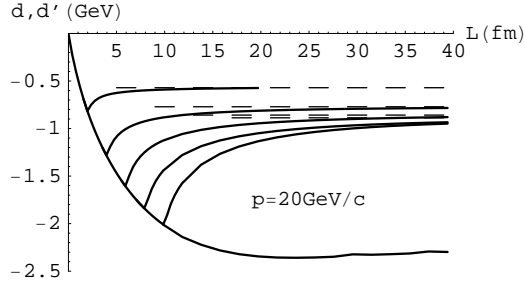}
\caption[*]{The function $d'(L,L_p)$ of \eq{twoterms} for $L_p =$ 2, 4, 6, 8 and $10\,\rm{fm}$ (from top to bottom) together with 
the function $d(L)$ (lower curve) of Fig.~\ref{dplot}, for $p = 20\,\rm{GeV}$. The dashed straight lines represent the asymptote
$d'_\infty$ of $d'(L,L_p)$ for the different values of $L_p$.}
\label{ddprime-fig} 
\end{figure}

The function $d'(L,L_p)$ defined in \eq{twoterms} is shown in Fig.~\ref{ddprime-fig}. For finite $L_p$, $d'(L,L_p)$ has a cusp at $L=L_p$, 
where the effect of the transition fields on the work done on the heavy quark starts\footnote{Strictly speaking, 
the effect of transition on $d'(L,L_p)$ starts at $L = L_s = 2v L_p/(1+v) < L_p$, in agreement with causality, 
as shown in Appendix B, see \eq{causality}. For $p \geq 10\,\rm{GeV}$, $v\simeq 1$ and the difference between $L_s$ and $L_p$ cannot be
seen on Fig.~\ref{ddprime-fig}.}. Most importantly, the asymptote $d'_\infty$ of $d'(L,L_p)$ in the limit 
$L \gg L_p \to \infty$ {\it scales in $\gamma$} when $\gamma \to \infty$ (see \eq{appdprimeinf}), 
\beq
d'_\infty \mathop{\ \ \simeq \ \ }_{v\to 1} \frac{1}{3} d_\infty \mathop{\ \ \simeq \ \ }_{v\to 1}  - \frac{1}{3}\sqrt{2}  
C_F \alpha_s m_D \gamma  \ \ .
\label{dprimeinf}
\eeq

As discussed in the end of section \ref{secinf}, in the case of a finite plasma size it is legitimate to 
interpret the {\it mechanical work} $d'(L = \infty, L_p)$ as quark {\it energy loss}. It is also clear that
$d'(\infty, L_p)$ includes the (radiative) contributions from initial bremsstrahlung and transition radiation. 
In order to obtain the 'zeroth order' loss $-\Delta E(L=\infty,L_p)\equiv -\Delta E_{0}(L_p)$, we must add to $d'(\infty,L_p)$ 
the stationary law $-\Delta E_{coll}(L_p)$ for purely collisional loss. For the latter we take 
the HTL result of \cite{BT} in the logarithmic accuracy $\log{(k_\mathrm{max}/m_g)} \gg 1$, 
\beqa
\label{linearlaw}
&& -\Delta E_{coll}(L_p) = \frac{C_F \alpha_s m_D^2 L_p}{2} \, \left[ \frac{1}{v}
- \frac{1-v^2}{2v^2}\log{\left( \frac{1+v}{1-v}\right) } \right]
\log{\left(\frac{k_\mathrm{max}}{m_g} \right)}\ ,  \\
&& {\rm where} \hskip .2cm k_\mathrm{max} \equiv {\rm Min \,}\left\{ \frac{ET}{M},
\sqrt{ET} \right\} \ \   {\rm and} \hskip .2cm  m_g = m_D/\sqrt{3}  \ \  .  \nn
\eeqa
The resulting energy loss
\beqa
-\Delta E(L=\infty, L_p) &=&  -\Delta E_{coll}(L_p)+ d'(L=\infty,L_p)  \nn \\
&=&  -\Delta E_{coll}(L_p)+ d(L_p) - W'(L=\infty,L_p)
\label{threeterms}
\eeqa
is determined by \eq{dofl}, \eq{linearlaw} and \eq{mastereq}, with the functions $h_1$ and $h_2$ given by \eq{h1h2app} 
and \eq{I1I2exp}. 

\begin{figure}[t]
\centering
\includegraphics[width=9cm]{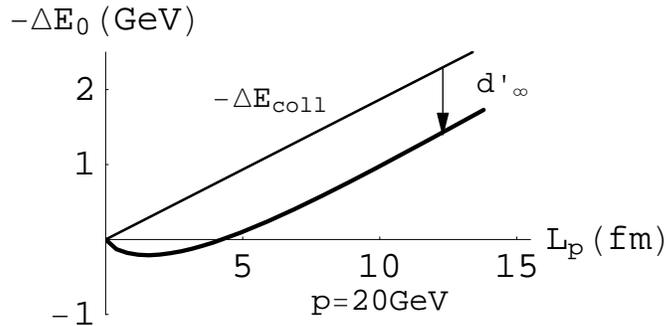}
\caption[*]{Charm quark energy loss $-\Delta E_0(L_p) = -\Delta E(L=\infty, L_p)$ for $p = 20\,\rm{GeV}$ (solid line) 
given by \eq{threeterms}.  $-\Delta E_0(L_p)$ is compared to the stationary law \eq{linearlaw} for collisional 
energy loss $-\Delta E_{coll}(L_p)$ (straight line) \cite{BT}.}
\label{DeltaEofLp-fig} 
\end{figure}

Our result for $-\Delta E_0(L_p)$ is shown in Fig~\ref{DeltaEofLp-fig}. We observe an important suppression
of the zeroth order energy loss as compared to the stationary collisional loss $-\Delta E_{coll}(L_p)$.
This is mainly due to the fact that the asymptote $d'_\infty$ of $d'(L=\infty, L_p)$, though reduced by a factor $1/3$ compared to 
$d_\infty$ (see \eq{dprimeinf}), still scales as $\gamma$ when $\gamma \gg 1$. As already mentioned in the 
end of section \ref{secinf}, this is due to a non-compensation between the TM effect and transition radiation. 
As is obvious from Fig~\ref{DeltaEofLp-fig}, we can define an {\it effective retardation time} $t_\mathrm{ret}$ before 
the linear regime for $-\Delta E_0(L_p)$ sets in, 
\beq
\label{tret}
t_\mathrm{ret} = d'_\infty/(dE/dx)_{coll}  \ \ ,
\eeq
leading to the scaling $t_\mathrm{ret} \propto \gamma/ m_D$, thus to a large effective delay.
Numerically, $t_{\rm ret} \simeq  5~{\rm fm}$ for $p = 20\,\rm{GeV}$. Fig~\ref{DeltaEofLp-fig} is our main result, which 
shows that invoking only the heavy quark collisional energy loss $-\Delta E_{coll}(L_p)$ 
to explain the 'single electron puzzle' is not satisfactory.
Indeed, adding the other relevant contributions (initial bremsstrahlung and transition radiation) to obtain the 
energy loss $-\Delta E_0(L_p)$ to zeroth order in an opacity expansion \cite{GLV}, 
we find that $-\Delta E_0(L_p) \ll -\Delta E_{coll}(L_p)$ for in-medium quark path length $L_p \lsim 5-7~{\rm fm}$.

\section{Conclusion}
\label{secconcl} 

In this paper we studied the energy loss $-\Delta E_0(L_p)$ of a fast heavy quark produced in a finite size QGP, to 
zeroth order in an opacity expansion. $-\Delta E_0(L_p)$ includes collisional energy loss, initial bremsstrahlung 
arising from the quark creation at $t=0$, and transition radiation appearing when the quark passes the discontinuity between 
medium and vacuum. Only the radiative loss {\it induced by rescatterings} in the plasma $-\Delta E_{rad}(L_p)$ 
\cite{Erad,GLV} should be added to $-\Delta E_0(L_p)$ to obtain the total heavy quark energy loss. 
Our main result is that $-\Delta E_0(L_p)$ is strongly reduced compared to the stationary linear law for collisional loss, as 
seen on Fig~\ref{DeltaEofLp-fig}. Due to the negative shift $d'_\infty$ scaling as $\gamma$, the linear regime for 
$-\Delta E_0(L_p)$ sets in after a quite large effective retardation time $t_{\rm ret} \propto \gamma$. 

From Fig.~\ref{DeltaEofLp-fig} we conclude that for the generic values $L_p \sim 5 \ {\rm fm}$ and 
$p = 20 \ {\rm GeV}$, the loss $-\Delta E_0(L_p)$ should be negligible, and the total heavy quark energy loss
should be theoretically correctly estimated by $-\Delta E_{rad}(L_p)$ only. Our results suggest that a missing contribution 
to the heavy quark energy loss, as proposed in Ref.~\cite{Wicks:2005gt}, 
might not be a correct explanation for the surprisingly strong nuclear attenuation of 
large $p_\perp$ electron spectra from heavy flavour decays.

We already mentioned in the Introduction that our results should be considered at the qualitative level, in particular because
the slope $(-dE/dx)_{coll}$ of the stationary regime usually considered \cite{TG,BT} is modified when the 
running of $\alpha_s$ is taken into account \cite{Peshier:2006hi}. 
As discussed in \cite{PGG}, the small $L$ (small $L_p$) behaviour of Fig.~\ref{dplot} (Fig.~\ref{DeltaEofLp-fig}) 
might also be affected by 
the running of $\alpha_s$. However the position and scaling in $\gamma$ 
of the asymptotes $d_\infty$ and $d'_\infty$ (see \eq{dprimeinf} and Fig.~\ref{DeltaEofLp-fig}) 
should be unaffected by the running of the coupling \cite{PGG}, leaving unchanged the main result of the present 
study. 

\vskip 1cm \acknowledgments  We wish to thank D.~Schiff, Y.~L.~Dokshitzer, and T.~Gousset for 
very stimulating exchanges during this work. We also thank A.~Adil and the authors of Ref.~\cite{Adil:2006ei}
for a useful correspondence, despite a disagreement on several crucial points. 


\appendix

\section{Derivation of the transition fields}

In this Appendix we obtain the functions $h_1(\omega, k_\perp)$ and 
$h_2(\omega, k_\perp)$ defining the transition fields \eq{homfields}. Written
in the mixed space $(\omega,\vec{k}_\perp,z)$, the continuity conditions at the surface $z=L_p$ read:
\beqa
\left[\vec{E}^{\rm in}_{1}(\omega,\vec{k}_\perp,L_p) + \vec{E}_{T1}^{\, '}(\omega,\vec{k}_\perp,L_p) \right]_\perp
= \left[\vec{E}^{\rm in}_{2}(\omega,\vec{k}_\perp,L_p) + \vec{E}_{T2}^{\, '}(\omega,\vec{k}_\perp,L_p) \right]_\perp \hskip 8mm && 
\\
\left[\vec{D}^{\rm in}_{1}(\omega,\vec{k}_\perp,L_p) + \epsilon_1 \vec{E}_{T1}^{\, '}(\omega,\vec{k}_\perp,L_p)\right]_z 
= \left[\vec{D}^{\rm in}_{2}(\omega,\vec{k}_\perp,L_p) + \epsilon_2 \vec{E}_{T2}^{\, '}(\omega,\vec{k}_\perp,L_p)\right]_z \hskip 5mm &&
\label{continuity}
\eeqa
where the color index is implicit and the inhomogeneous fields are given by ($i = 1,2$)
\beqa
\vec{E}^{\rm in}_{i}(\omega,\vec{k}_\perp,z)&=&
\int \frac{dk^z}{2\pi}\, e^{i k^z z} \left[ \vec{E}_{Li}
(\omega,\vec{k}_\perp,k^z)+
 \vec{E}_{Ti}(\omega,\vec{k}_\perp,k^z)\right] \\
\vec{D}^{\rm in}_{i}(\omega,\vec{k}_\perp,z)&=&
\epsilon_{i}(\omega) \vec{E}^{\rm in}_{i}(\omega,\vec{k}_\perp,z) \ \ \ .
\eeqa
Using the definition \eq{homfields}, we can easily solve for the functions 
$h_1$ and $h_2$. We find from the above continuity conditions:
\beqa
h_1(\omega, k_\perp) &=& \frac{\epsilon_2 - \epsilon_1}{\epsilon_1 \sigma_2+ \epsilon_2 \sigma_1} 
\left[\epsilon_2  J + \sigma_2 K \right]  \nn \\
h_2(\omega, k_\perp) &=& \frac{\epsilon_1 - \epsilon_2}{\epsilon_1 \sigma_2+ \epsilon_2 \sigma_1} 
\left[- \epsilon_1 J + \sigma_1 K \right]  \ \ ,
\label{h1h20}
\eeqa
with 
\begin{equation}
J=\frac{q^a \vec{k}_\perp \cdot \left[\vec{E}^{{\rm in} \,a}_2(L_p)- \vec{E}^{{\rm in} \,a}_1(L_p)\right]_\perp}
{4\pi C_F \alpha_s k_\perp^2 (\epsilon_2-\epsilon_1)}
\quad{\rm and}\quad
K=\frac{q^a  \left[\vec{D}^{{\rm in} \,a}_2(L_p)- \vec{D}^{{\rm in} \,a}_1(L_p)\right]_z}
{4\pi C_F \alpha_s k_\perp^2 (\epsilon_2-\epsilon_1)} \ \ \ .
\end{equation}
Using \eq{linappr} (with $\epsilon_{Li}=\epsilon_{Ti}=\epsilon_i$) and \eq{jmom} we obtain:
\beqa
J(\omega, k_\perp) 
&=&  \frac{v}{\omega \epsilon_1 \epsilon_2 } \int \frac{dk^z}{2\pi} 
\frac{e^{i k^z L_p} \, k^z \, \left[(\epsilon_1 + \epsilon_2) \omega^2 - k^2\right]}{(\omega-v k^z) (\epsilon_1 \omega^2 - k^2)(\epsilon_2 \omega^2 - k^2)} \nn \\ 
K(\omega, k_\perp) &=& - v \omega \int \frac{dk^z}{2\pi} 
\frac{e^{i k^z L_p}}{(\omega-v k^z) (\epsilon_1 \omega^2 - k^2)(\epsilon_2 \omega^2 - k^2)} \ \ .
\label{JK}
\eeqa
From \eq{h1h20} and \eq{JK} we readily check that $h_i(-\omega, k_\perp)=
h_i(\omega, k_\perp)^*$. Hence the transition fields \eq{homfields} satisfy 
$\vec{E}_{T}^{\, '}(-\omega,-\vec{k}_\perp,z) = 
\vec{E}_{T}^{\, '}(\omega,\vec{k}_\perp,z)^*$, ensuring the reality of the 
fields in coordinate space. We can further put \eq{h1h20} in the form 
\beqa
h_1(\omega, k_\perp) &=& i  \frac{(\epsilon_2 - \epsilon_1) \omega}{\epsilon_1 \sigma_2+\epsilon_2 
\sigma_1}  \, I_1 \nn \\
h_2(\omega, k_\perp) &=& i  \frac{(\epsilon_1 - \epsilon_2) \omega}{\epsilon_1 \sigma_2+\epsilon_2 
\sigma_1}\,I_2\ 
\ ,
\label{h1h2app}
\eeqa
where the functions $I_1$ and $I_2$ read
\beqa
I_1 &\equiv&  - \frac{i v}{\epsilon_1 \omega^2} \int \frac{dk^z}{2\pi} 
\frac{e^{i k^z L_p}}{(\omega-v k^z) (\epsilon_1 \omega^2 - k^2)} \left[\frac{k^z(k^z+\sigma_2)-\epsilon_1 \omega^2}{k^z+\sigma_2}\right]
\nn \\
I_2 &\equiv&  - \frac{i v}{\epsilon_2 \omega^2} \int \frac{dk^z}{2\pi} 
\frac{e^{i k^z L_p}}{(\omega-v k^z)(\epsilon_2 \omega^2 - k^2)} \left[ \frac{k^z(\sigma_1 -k^z)+\epsilon_2 \omega^2}{k^z-\sigma_1}\right] 
\label{I1I2}
\eeqa 
We calculate $I_1$ and $I_2$ by closing the contour of the $k^z$-integral in the upper half of the complex plane. We obtain:
\beqa
I_1 &=& \frac{1}{\epsilon_1 \omega^2} \left\{ \frac{e^{i\frac{\omega}{v} L_p}(\sigma_2 \frac{\omega}{v} -\epsilon_1 \omega^2 + \frac{\omega^2}{v^2}  )}{ \left[ \frac{\omega^2}{v^2} -\sigma_1^2 \right] \left[ \frac{\omega}{v} +\sigma_2 \right]} 
+  \frac{e^{i\sigma_1 L_p}(\sigma_1 \sigma_2 - k_\perp^2)}{2 \sigma_1 \left[\sigma_1 - \frac{\omega}{v}\right] 
\left( \sigma_1 + \sigma_2\right)}  \right\}  \nn \\
I_2 &=& \frac{1}{\epsilon_2 \omega^2} \left\{ \frac{e^{i\frac{\omega}{v} L_p}(\sigma_1 \frac{\omega}{v} +\epsilon_2 \omega^2 - \frac{\omega^2}{v^2})}{\left[ \frac{\omega^2}{v^2} -\sigma_2^2 \right] \left[ \frac{\omega}{v} -\sigma_1 \right]} 
+ \frac{e^{i\sigma_2  L_p}(\sigma_1 \sigma_2 + k_\perp^2)}{2 \sigma_2 \left[ \sigma_2 - \frac{\omega}{v}\right] 
\left( \sigma_2 - \sigma_1 \right)} +  \frac{\epsilon_2\, e^{i\sigma_1  L_p}}{(\epsilon_1 - \epsilon_2) \left[\sigma_1- \frac{\omega}{v}\right]}
\right\}  \nn \\
\label{I1I2exp}
\eeqa
The transition fields \eq{homfields} are determined by \eq{h1h2app} and \eq{I1I2exp}, where we recall that 
the retarded prescription $\omega \to \omega + i\eta$ is implicit, and that the square root satisfies 
$\im \sqrt{z} \geq 0$ for complex $z$.

\section{Some properties of $-W'(L,L_p)$}

Here we study some features of the contribution $-W'(L,L_p)$ to the heavy quark energy loss 
(see \eq{mastereq} and \eq{threeterms}) which arises from the transition fields \eq{homfields}. 
We first show that this contribution is consistent with causality. Then we derive, for $\gamma \gg 1$, the limit 
of $-W'(L,L_p)$ for $L \gg L_p \to \infty$. This limit corresponds to the amount to be added to the
asymptotic value $d_\infty$ of $d(L)$ to reach the asymptote $d'_\infty$ of $d'(L)$ (see Fig.~\ref{ddprime-fig}). 

\centerline{\bf Causality}
\vskip 5mm 

From causality the heavy quark cannot be sensitive to the transition fields right after its production at
$t=0$. The effect of transition fields can start only when a signal emitted by the quark at $t=0$ (and $z=0$) along
the positive $z$-axis has had time to reach the medium-vacuum separation surface located at $z=L_p$, and to come back 
to the position of the quark. Since the quark is moving with velocity $v$, and the largest speed of light in a medium of 
permittivity \eq{model} is unity (corresponding to $\omega \to \infty$), this time is given by 
$t_s = L_s/v= 2 L_p -L_s$. Thus $-W'(L,L_p)$ must vanish when the quark has travelled less than the distance $L_s$, 
\beq
L < L_s = \frac{2v L_p}{1+v}  \Rightarrow -W'(L,L_p) = 0  \ \ .
\label{causality}
\eeq

The latter causality requirement can be proved directly from \eq{mastereq}. For $L < L_p$ only the first term in the 
bracket of \eq{mastereq} contributes. Inserting the expression of $h_1$ given by \eq{h1h2app} and \eq{I1I2exp} we obtain:
\beqa
\left. -W'(L, L_p) \right|_{L<L_p}= \frac{C_F \alpha_s}{2 \pi} \int_0^{\infty} dk_\perp^2 \, k_\perp^2 \int_{-\infty}^{\infty} 
\frac{d\omega}{\omega} 
\frac{e^{i(\frac{\omega}{v} +\sigma_1) L_p}-e^{i(\frac{\omega}{v} +\sigma_1)(L_p-L)}}{\frac{\omega}{v} +\sigma_1} && \nn \\ 
\times \frac{1-\epsilon_2/\epsilon_1}{\epsilon_1 \sigma_2+\epsilon_2 \sigma_1}  \,
\left\{ \frac{\sigma_2 \frac{\omega}{v} -\epsilon_1 \omega^2 + \frac{\omega^2}{v^2}}{ \left[ \frac{\omega^2}{v^2} -\sigma_1^2 \right] \left[ \frac{\omega}{v} +\sigma_2 \right]} +  \frac{e^{i(\sigma_1 -\frac{\omega}{v}) L_p}(\sigma_1 \sigma_2 - k_\perp^2)}{2 \sigma_1 \left[\sigma_1 - \frac{\omega}{v}\right] 
\left( \sigma_1 + \sigma_2\right)}  \right\} \ \ . \ \ \ \ \ &&
\label{caus1}
\eeqa
Recalling (see section 3) that the square root satisfies 
$\im \sqrt{z} \geq 0$ for complex $z$, we can perform the $\omega$-integral by closing the integration contour in the upper half-plane
provided the phase 
\beq
2 \sigma_1 L_p - \left( \frac{\omega}{v} +\sigma_1 \right) L 
\eeq
has a positive imaginary part when $|\omega| \to \infty$ in this half-plane. 
Since for $\im \omega > 0$ the integrand of \eq{caus1} has no singularity (which can be easily checked) and 
$\sigma_1 \simeq \omega$ when $|\omega| \to \infty$, we obtain that $-W'(L, L_p)$ vanishes when 
$2 L_p > (1+1/v)L$, which proves \eq{causality}. 

\vskip 5mm 
\centerline{\bf Limit of $-W'(L,L_p)$ when $L \gg L_p \to \infty$}
\vskip 5mm 

We first consider the limit of $-W'(L,L_p)$ for $L \to \infty$ at fixed $L_p$. 
This represents the amount brought by transition radiation, to be added to 
$-\Delta E_{coll}(L_p)+d(L_p)$ (see \eq{threeterms})
in order to get the heavy quark energy loss $-\Delta E_0(L_p)$ represented in 
Fig.~\ref{DeltaEofLp-fig}. For a fast quark, $\gamma \gg 1$, this limit is given by the second term 
of \eq{mastereq}, where the term with a rapidly oscillating phase factor $\propto e^{-i(\frac{\omega}{v} -\sigma_2) L}$ 
is neglected:
\beq
-W'(L\to \infty ,L_p) =  \frac{C_F \alpha_s}{2 \pi} \int_0^{\infty} dk_\perp^2 \, k_\perp^2 
\int_{-\infty}^{\infty} d\omega \, \frac{e^{-i \frac{\omega}{v} L_p} (\epsilon_2 - \epsilon_1) \omega}{\left[ \frac{\omega}{v} -\sigma_2 \right] (\epsilon_1 \sigma_2+\epsilon_2 \sigma_1)} \, \, I_2 \ \ ,
\label{Ltoinf}
\eeq
where we used \eq{h1h2app} and $I_2$ is given in \eq{I1I2exp}. 
 
We now evaluate the limit of \eq{Ltoinf} when $L_p \to \infty$. In this limit 
only the first term of $I_2$ in \eq{I1I2exp} contributes to \eq{Ltoinf}, the two other terms oscillating rapidly. We get:
\beqa
\left. -W'(L ,L_p) \right|_{L \gg L_p \to \infty} = \hskip 4cm && \nn \\
= \frac{C_F \alpha_s}{2 \pi} \int_0^{\infty} dk_\perp^2 \, k_\perp^2 
\int_{-\infty}^{\infty} \frac{d\omega}{\omega} 
\frac{(1 - \epsilon_1/\epsilon_2)\left(\sigma_1 \frac{\omega}{v} +\epsilon_2 \omega^2 - \frac{\omega^2}{v^2}\right)}{ (\epsilon_1 \sigma_2+\epsilon_2 \sigma_1)\left[ \frac{\omega}{v} +\sigma_2 \right] \left[\frac{\omega}{v} -\sigma_2 \right]^2 \left[ \frac{\omega}{v} -\sigma_1 \right]}  \ \ . && 
\label{LLptoinf2}
\eeqa
The latter integral can be evaluated for $\gamma \gg 1$ by anticipating that it arises dominantly from the 
domain $\omega \sim \gamma m$ and $k_\perp \sim m$. 
Approximating the integrand in this region and using \eq{model} (and $\epsilon_2 =1$) we obtain
\beqa
\left. -W'(L ,L_p) \right|_{L \gg L_p \to \infty} 
\mathop{\ \ \simeq \ \ }_{v\to 1}\frac{m^2 C_F \alpha_s}{\pi} \int_0^{\infty} dk_\perp^2 \, k_\perp^2 
\int_{-\infty}^{\infty} \frac{d\omega}{(\frac{\omega^2}{\gamma^2} + k_\perp^2 )^2 (\frac{\omega^2}{\gamma^2} + k_\perp^2 + m^2)} \ \ . \nn \\ 
\label{LLptoinf3}
\eeqa
The latter double integral can be exactly evaluated, yielding (use $m= m_D/\sqrt{2}$):
\beqa
\left. -W'(L ,L_p) \right|_{L \gg L_p \to \infty} \equiv d'_\infty - d_\infty \mathop{\ \ \simeq \ \ }_{v\to 1}
\frac{2}{3} \sqrt{2} C_F \alpha_s m_D \gamma \ \ .
\label{LLptoinf4}
\eeqa
The expression \eq{LLptoinf3} is also 
trivially shown to be dominated by $\omega \sim \gamma m$, $k_\perp \sim m$, justifying our initial approximation. 
The result \eq{LLptoinf4} is exactly $2/3$ of $|d_\infty|$, see \eq{dinf2}. Hence:
\beq
d'_\infty \mathop{\ \ \simeq \ \ }_{v\to 1} \frac{1}{3} d_\infty \mathop{\ \ \simeq \ \ }_{v\to 1}  - \frac{1}{3}\sqrt{2}  C_F \alpha_s m_D \gamma  \ \ .
\label{appdprimeinf}
\eeq

Finally, we note that in \eq{I1I2exp} the second and third terms of $I_2$, which have been neglected to extract the
large $L_p$ limit of \eq{Ltoinf}, are proportional to the phase factors  
$e^{-i(\frac{\omega}{v} -\sigma_{1,2}) L_p}$. When $\omega \sim \gamma m \gg m$ and $k_\perp \sim m$ we can approximate, 
for instance, 
\beq
\left( \frac{\omega}{v} -\sigma_1 \right) L_p 
\simeq \frac{L_p}{2\omega} \left( \frac{\omega^2}{\gamma^2} + k_\perp^2 + m^2\right) \sim \morder{L_p m/\gamma} \ \ , 
\eeq
Thus the large $L_p$ limit used above should be understood as $L_p \gg \gamma/m_D$, where the neglected terms are
indeed rapidly oscillating. In this limit the function $d'(L,L_p)$ is close to its asymptote $d'_\infty$, \ie, 
the energy loss $-\Delta E_0(L_p)$ including the effect of 
transition radiation (see Fig.~\ref{DeltaEofLp-fig}) is close to its asymptotic regime. 

\providecommand{\href}[2]{#2}\begingroup\raggedright
\endgroup

\end{document}